\documentclass{aa}
\usepackage{txfonts}
\usepackage{hyperref}
\makeatletter
\renewcommand*\aa@pageof{, page \thepage{} of \pageref*{LastPage}}
\makeatother
\usepackage[T1]{fontenc}

\DeclareRobustCommand{\VAN}[3]{#2}
\let\VANthebibliography\thebibliography
\def\thebibliography{\DeclareRobustCommand{\VAN}[3]{##3}\VANthebibliography}

\usepackage{graphicx}	
\usepackage{amsmath}	
\usepackage[normalem]{ulem}
\graphicspath{ {./figs/} }

\usepackage[free-standing-units = true,
space-before-unit = true,
use-xspace = true,
per-mode = power-positive-first]{siunitx}
\usepackage{xcolor}

\hypersetup{
	colorlinks,
	linkcolor={red!50!black},
	citecolor={blue!50!black},
	urlcolor={blue!80!black}
}

\DeclareSIUnit \msun {\text{\ensuremath{M_\odot}}}
\DeclareSIUnit \rsun {\text{\ensuremath{R_\odot}}}
\DeclareSIUnit \lsun {L_\odot}

\DeclareSIUnit \year {yr}
\DeclareSIUnit \kms {\kilo \meter \per\second}

\newcommand{\GG}[1]{}
\definecolor{barbiepink}{rgb}{0.8784 0.1294 0.5412}

\begin{document}

   \title{Population synthesis of Thorne-\.Zytkow objects}
    \subtitle{Rejuvenated donors and unexplored progenitors in the common envelope formation channel}

   \author{K. Nathaniel\inst{1,2}\thanks{\email{kjn9240@rit.edu}}
          \and
          A. Vigna-G\'omez\inst{1}\thanks{\email{avigna@mpa-garching.mpg.de}}
          \and
          A. Grichener\inst{1,3,4}
          \and
          R. Farmer\inst{1}
          \and
          M. Renzo\inst{3}
          \and
          R. W. Everson\inst{5,6}
          }

   \institute{Max-Planck-Institut f\"ur Astrophysik, Karl-Schwarzschild-Str. 1, D-85748 Garching, Germany 
         \and
        Center for Computational Relativity and Gravitation, Rochester Institute of Technology, 85 Lomb Memorial Drive, Rochester, NY 14623, USA  
        \and
        Steward Observatory, University of Arizona, 933 North Cherry Avenue, Tucson, AZ 85721, USA
        \and
        Department of Physics, Technion, Haifa, 3200003, Israel
        \and
        Department of Astronomy \& Astrophysics, University of California, Santa Cruz, CA 95064, USA
        \and
        Department of Physics and Astronomy, University of North Carolina at Chapel Hill, 120 E. Cameron Ave, Chapel Hill, NC 27599, USA
             }


 \abstract{}{}{}{}{}
 
  \abstract
   { 
   Common envelope evolution of a massive star and a neutron star companion has two possible outcomes: formation of a short-period binary (a potential gravitational wave source progenitor) or a merger of the massive star with the neutron star. 
   If the binary merges, a structure with a neutron star core surrounded by a large diffuse envelope, a so-called Thorne-\.Zytkow object (T\.ZO), may form.
   The predicted appearance of this hypothetical class of star is very similar to red supergiants, making observational identification difficult.
    }
   {Our objective is to understand the properties of systems that are potential T\.ZO progenitors, particularly binary systems that enter a common envelope phase with a neutron star companion. We also aim to distinguish those that have been through a previous stable mass transfer phase, which can rejuvenate the accretor.
} 
   {We use the rapid population synthesis code COMPAS at Solar metallicity and with common envelope efficiency parameter set to unity to determine the population demographics of T\.ZOs.
   We use one-dimensional evolutionary T\.ZO models from the literature to determine a fit for T\.ZO lifetime in order to estimate the current number of T\.ZOs in the Milky Way as well as to assess core disruption during the merger.
   }
   {
   We explore the progenitors in the Hertzsprung-Russell diagram, calculate formation rates, and investigate kinematics of the progenitor stars.
   We find that the vast majority ($\approx$\qty{92}{\percent}) of T\.ZO progenitors in our population have experienced mass transfer and become rejuvenated before their formation event. 
   In the Milky Way we estimate a T\.ZO formation rate of \qty{\approx 4e-4}{\per \year}, which results in \num{\approx 5\pm 1} T\.ZOs at present.  
   }
    {}
\titlerunning{Population synthesis of T\.{Z}Os}
\authorrunning{K. Nathaniel et al.}
   \maketitle
%

\nolinenumbers
\section{Introduction}\label{sec:intro}

In the late 1930s, \citet{landauOriginStellarEnergy1938} and \citet{oppenheimerMassiveNeutronCores1939} first theorized that stars could have dense neutron cores. Four decades later, \citet{thorneRedGiantsSupergiants1975, thorneStarsDegenerateNeutron1977} predicted and calculated the first structural models of a stable configuration of a star with a neutron star (NS) core surrounded by a diffuse envelope, later designated as a Thorne-\.Zytkow object (T\.ZO).

T\.ZOs are a theoretical class of star that appear like an ordinary M supergiant on the surface, despite the NS core hiding underneath (e.g., \citealt{thorneStarsDegenerateNeutron1977, farmerObservationalPredictionsThorne2023}).
There are multiple proposed formation channels: the merger of a non-degenerate star and NS via common envelope (CE) evolution
(e.g., \citealt{taamDoubleCoreEvolution1978}, \citealt{termanDoubleCoreEvolution1995}, but see also \citealt{ablimitStellarCoremergerinducedCollapse2022} where T\.ZOs are proposed to form via CE evolution with a white dwarf core),
direct collision of the NS with the stellar core via a kick \citep[e.g.,][]{leonardNewWayMake1994,hiraiNeutronStarsColliding2022}, or dynamical mergers within a cluster or triple system \citep[e.g.,][]{rayEvolutionStellarBinaries1987,eisnerPlanetHuntersTESS2022}. The collision and dynamical formation channels are thought to have very low rates comparatively \citep{podsiadlowskiEvolutionFinalFate1995, renzoMassiveRunawayWalkaway2019,grichenerMergersNeutronStars2023}, thus the focus of this paper is the CE channel.

\begin{figure}[!htbp]
	\centering
	\includegraphics[width=\columnwidth]{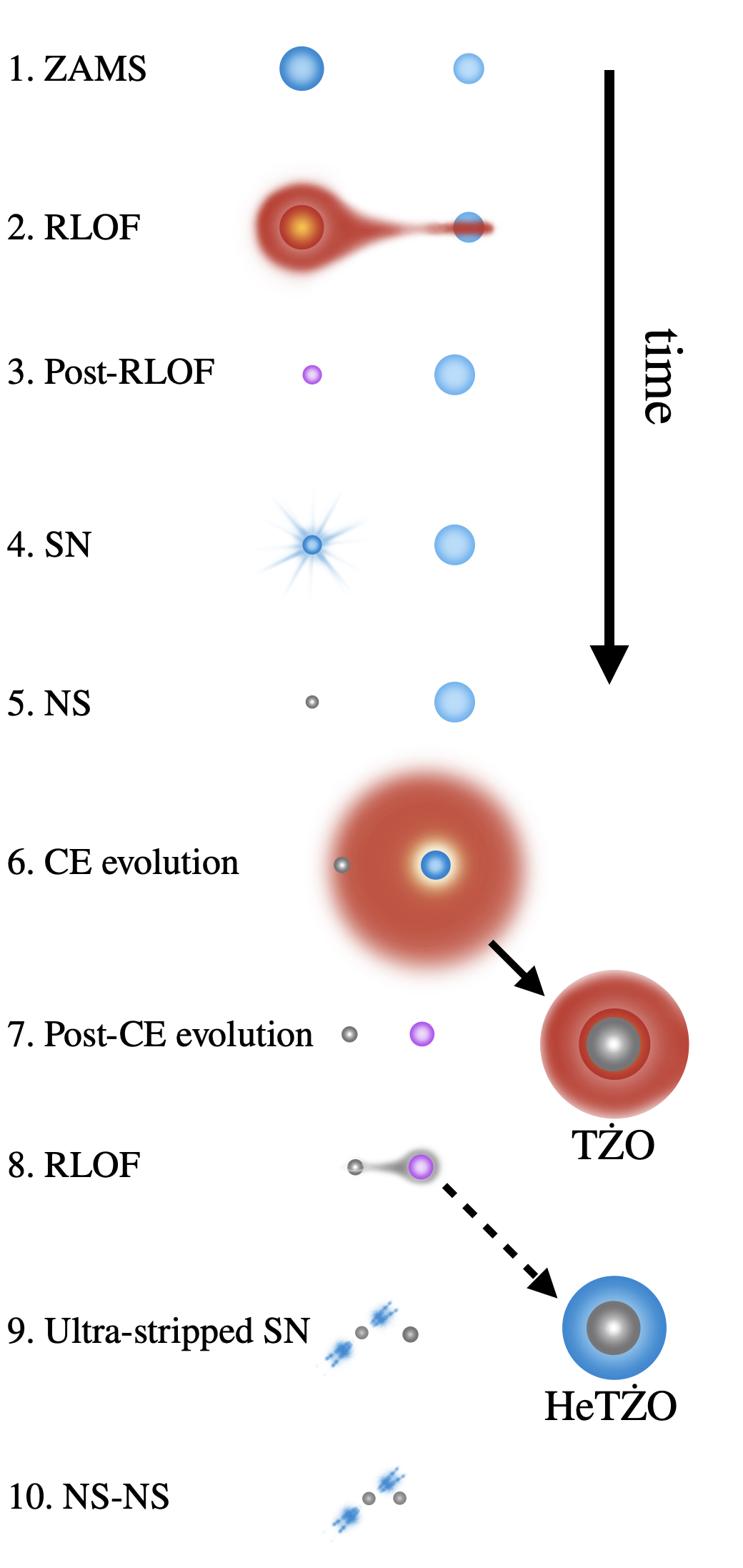}
	\caption{Schematic showing the CE formation channel for T\.ZOs, based on Fig.~1 of \cite{vigna-gomezCommonEnvelopeEpisodes2020,vigna-gomezGraphicsEvolutionaryPathways2020}. The chronology and the nomenclature is the following. 1. Zero-age main sequence (ZAMS); 2. Roche lobe overflow (RLOF); 3. Post-RLOF, or envelope detachment; 4. Supernova (SN) of the primary; 5. Neutron star (NS) forms from the SN explosion; 6. Common envelope evolution (CE evolution); 7. Post-CE evolution, which can lead to either the ejection of the envelope, forming a stripped star + NS, or the formation of a T\.ZO from the merger of the NS and the stellar core; 8. RLOF of the stripped binary, further stripping the companion star; 9. Ultra-stripped SN of the companion star; 10. NS binary. The solid arrow from step 6 indicates the CE formation channel for T\.ZOs and the dashed arrow from step 8 shows the alternative channel to a T\.ZO with a HeMS donor star.
    }
	\label{fig:evol_channel}
\end{figure}

T\.ZOs forming via CE evolution share a common channel with most NS binaries (NS-NS) in the Galaxy \citep{taurisFormationDoubleNeutron2017}.
The channels are the same up to the formation of the CE; if the CE fails to eject a T\.ZO may form, otherwise a NS-NS forms with successful envelope ejection.
This same channel was suggested to account for GW170817 (e.g., \citealt{vigna-gomezFormationHistoryGalactic2018,kruckowProgenitorsGravitationalWave2018}), the first gravitational wave signal observed by the Laser Interferometer Gravitational-wave Observatory (LIGO) and Virgo from the inspiral of a NS-NS \citep{abbottGW170817ObservationGravitational2017}.
An outline of the CE channel is shown in Fig.~\ref{fig:evol_channel}, starting with the binary at the zero-age main sequence (ZAMS) in step 1.
After undergoing mass transfer (MT) through Roche lobe overflow (RLOF) and stripping (steps 2 and 3), the first supernova (SN) occurs (step 4).
At this point, when there is just one NS and one non-degenerate star (step 5), the binary has several possible outcomes.
If the explosion ejects enough mass, the natal kick received by the NS is high enough, or both, the binary may become unbound \citep[e.g.,][]{blaauwOriginBtypeStars1961,boersmaMathematicalTheoryTwobody1961} --such is the fate of the vast majority of massive binaries \citep[e.g.,][]{eldridgeRunawayStarsProgenitors2011,renzoMassiveRunawayWalkaway2019}. If the binary remains bound and the orbital evolution imposes it, a CE may follow (step 6; e.g., \citealt{paczynskiCommonEnvelopeBinaries1976,taamDoubleCoreEvolution1978,webbinkDoubleWhiteDwarfs1984,ibenCommonEnvelopesBinary1993,ivanovaCommonEnvelopeEvolution2013,ivanovaCommonEnvelopeEvolution2020,ropkeSimulationsCommonenvelopeEvolution2023}), which is thought to be crucial to the formation of NS-NSs \citep[e.g.,][]{taurisFormationDoubleNeutron2017,vigna-gomezFormationHistoryGalactic2018,gallegos-garciaEvolutionaryOriginsBinary2023}.

During the CE phase, a non-degenerate star overflows its Roche lobe and begins dynamically unstable MT with its NS companion.
A CE forms around the NS and the core of the companion.
Drag and frictional forces in this configuration lead to a rapid inspiral phase that decreases the orbital separation of the binary.
If the envelope is ejected before the NS reaches the core, then the binary becomes a short-period NS + stripped-star binary \citep[e.g.][]{gotbergStarsStrippedBinaries2020}, the ideal candidate for a NS-NS progenitor (steps 7 through 10 in Fig.~\ref{fig:evol_channel}).
Alternatively, the NS and core can merge, resulting either in a luminous transient (e.g., \citealt{chevalierCommonEnvelopeEvolution2012,sokerExplainingIPTF14hlsCommonenvelope2018,sokerDiversityCommonEnvelope2019,dongTransientRadioSource2021,grichenerCommonEnvelopeJets2023}) or a T\.{Z}O. In this study, we use the terms ``T\.ZO'' and ``T\.ZO candidate'' interchangeably to describe these merger products, while remaining agnostic to the details of establishing a stable post-merger structure.

Following the pioneering work on structural models by \cite{thorneRedGiantsSupergiants1975,thorneStarsDegenerateNeutron1977}, \citet{biehleHighMassStarsDegenerate1991} extended the static model space by constructing models that had a convective burning region based on the rapid-proton process. The first evolutionary models of T\.ZOs were created by \citet{cannonStructureEvolutionThorneZytkow1992}, and \citet{cannonMassiveThorneZytkowObjects1993} and more recently revisited by \citet{farmerObservationalPredictionsThorne2023}.
High mass T\.ZOs ($\gtrsim \qty{12}{\msun}$) have a completely convective envelope and extremely high temperatures close to the NS \citep{eichGiantSupergiantStars1989,cannonMassiveThorneZytkowObjects1993}. Consequently, they undergo the interrupted rapid-proton (irp) process, which is expected to result in a unique chemical abundance profile \citep{biehleObservationalProspectsMassive1994,cannonMassiveThorneZytkowObjects1993} that may possibly be used as an observational signature. Lower mass T\.ZOs are supported by accretion onto the NS and are not expected to have a discernible observational signature from red giants; low mass T\.ZOs have a
radiative envelope and therefore no irp-process. For a deeper discussion of T\.ZO structure and chemical profile analysis, see \citet{cannonStructureEvolutionThorneZytkow1992}, \citet{cannonMassiveThorneZytkowObjects1993}, \citet{levesqueDiscoveryThorneZytkowObject2014}, and \citet{ogradyCoolLuminousHighly2020,ogradyCoolLuminousHighly2023,ogradyThorneZytkowObjects2024}.

T\.ZOs are an exciting prospect in stellar physics, as they are part of an exotic class of theoretical star-like objects.
Their interiors and unique nucleosynthetic processes are an excellent test of the capabilities of state-of-the-art stellar evolution codes \citep{levesqueDiscoveryThorneZytkowObject2014,farmerObservationalPredictionsThorne2023}.
They are excellent candidates for multimessenger astronomy, as the asymmetries and rapid rotation of the NS core are expected to emit continuous gravitational waves \citep{demarchiProspectsMultimessengerObservations2021}.
In practice, T\.ZOs are very difficult to distinguish from red (super)giants through electromagnetic only observations \citep{biehleHighMassStarsDegenerate1991,biehleObservationalProspectsMassive1994,ogradyCoolLuminousHighly2020,ogradyCoolLuminousHighly2023}.
The current candidates, HV 2112 \citep{levesqueDiscoveryThorneZytkowObject2014}, VX Sgr \citep{taberneroNatureVXSagitarii2021}, and HV 11417 \citep{beasorCriticalReevaluationThorneZytkow2018}, have all been identified through unusual chemical abundance profiles, although there is still much debate about their classification (e.g., \citealt{coeDeepInfraredOptical1998,vantureAquariiThorneZytkowObject1999,toutHV2112ThorneZytkowObject2014,maccaroneLargeProperMotion2016,taberneroNatureVXSagitarii2021,farmerObservationalPredictionsThorne2023}). 

In this study, we investigate T\.ZO progenitor systems in detail.
Though there are many processes in binary evolution that have been studied previously or merit further investigation, for simplicity we focus on one important aspect: the role of rejuvenation, which can change the stellar structure \citep{hellingsPostRLOFStructureSecondary1984, renzoRejuvenatedAccretorsHave2023}.
Rejuvenation occurs when a star in a binary experiences stable RLOF and transfers mass onto its main sequence (MS) companion (e.g., \citealt{neoEffectRapidMass1977}, \citealt{hellingsPhenomenologicalStudyMassive1983, hellingsPostRLOFStructureSecondary1984}, see step 2 of Fig.~\ref{fig:evol_channel}). Accretion causes the MS accretor to increase its mass, 
and consequently leads to an increase in the size of the convective core.
As the core mass increases, unburnt material from outside the core is mixed through the convective boundary, refueling the core with fresh hydrogen and rejuvenating the accretor star.
This process is not driven by rotation, as even non-rotating models can become rejuvenated \citep{renzoRejuvenatedAccretorsHave2023,waggAsteroseismicImprintsMass2024}.

The change in structure in the core-envelope boundary region due to rejuvenation affects the binding energy of the accretor after the MT phase. The core-envelope boundary region contains most of the mass of the envelope, therefore structure changes in that layer have a long-lasting impact on the envelope, which is amplified later on by the evolutionary contraction of the core. \citet{renzoRejuvenatedAccretorsHave2023} show that the binding energy at a fixed total mass and radius is still lower after rejuvenation than that of an unrejuvenated star of the same mass and radius, implying that if a binary system consisting of a NS and a rejuvenated star engages in a CE phase, it is possible that the less bound envelope of the rejuvenated star could be ejected and result in a short-period stripped binary instead of a T\.ZO. Most massive binaries leading to NS-NSs experience stellar rejuvenation \citep{vigna-gomezFormationHistoryGalactic2018,vigna-gomezCommonEnvelopeEpisodes2020}.
However, the impact of rejuvenation is not yet fully understood, and it is conceivable that it might have contradictory effects on the formation rate of T\.ZOs.

In this paper, we use rapid population synthesis in order to quantify and characterize binary systems that enter a CE phase with a NS companion.
We then distinguish between those that merge (i.e., T\.ZO candidates) and those that do not (i.e., NS-NS candidates).
We determine the potential impact of rejuvenation of the companion at the population level for NSs interacting with non-degenerate stars.
We calculate the formation rate of T\.ZOs per unit solar mass (\unit{\msun}) and per SN and estimate the predicted number of T\.ZOs in the Galaxy, as well as explore the potential effect of rejuvenation on the formation and merger rate of double compact objects (DCOs).
In Sect.~\ref{sec:synthpop} we discuss our synthetic population and how we identify and quantify systems of interest. In Sect.~\ref{sec:results} we present our results. We place them in context and discuss their broader meaning in Sect.~\ref{sec:discuss}. We finalize with a brief summary in Sect.~\ref{sec:summary}.

\section{Methods} \label{sec:synthpop}

\subsection{Data and initial distributions}\label{subsec:data_initialdists}

We use the synthetic stellar populations from \citet{grichenerMergersNeutronStars2023}\footnote[1]{Data publicly available via Zenodo (DOI 10.5281/zenodo.11237180, \citealt{grichenerPopulationSynthesisData2024}).}, generated with the rapid binary population synthesis code COMPAS\footnote[2]{\url{https://compas.science/}}, version 02.31.06 \citep{stevensonFormationFirstThree2017,vigna-gomezFormationHistoryGalactic2018,team-compasCOMPASRapidBinary2022}.
For emulating single-star evolution, COMPAS uses the analytical fits of \citet{hurleyComprehensiveAnalyticFormulae2000}, based on the stellar models of \citet{polsStellarEvolutionModels1998}; for binary evolution, COMPAS follows \citet{hurleyEvolutionBinaryStars2002}.
The stellar models extend up to \qty{50}{\msun}, therefore COMPAS extrapolates for stars of higher mass.
During binary evolution, COMPAS accounts for mass loss through stellar winds, MT via RLOF, CE episodes, SNe, and gravitational wave radiation. 
COMPAS version 02.31.06 does not account for rotation, magnetic braking, nor tidal interactions.
For more details, we refer to the COMPAS methods paper \citep{team-compasCOMPASRapidBinary2022}.

\citet{grichenerMergersNeutronStars2023} studies NSs and black holes (BHs) that merged with giant secondaries during CE evolution using synthetic populations of $10$ million massive binaries at various metallicities and CE efficiency parameters (see Eqn.~\ref{eqn:eformalist}).
In this study, we use the population created with solar metallicity ($Z = Z_\odot = 0.0142$, \citealt{asplundChemicalCompositionSun2009}) and CE efficiency parameter set to unity.
The setup is based on the \texttt{Fiducial} model from \citet{vigna-gomezFormationHistoryGalactic2018}.
We define the star with the larger ZAMS mass as the primary and denote its mass by $M_1$ throughout the evolution. In this synthetic population, the primary mass at the ZAMS is drawn using a \citet{kroupaVariationInitialMass2001} initial mass function (IMF) in the form $\mathrm{d}N/\mathrm{d}M_{1,\mathrm{ZAMS}} \propto M^{-2.3}_{1,\mathrm{ZAMS}}$ from \qtyrange{5}{100}{\msun}.
The mass $M_2$ of the secondary star is then calculated from a flat distribution in the initial mass ratio ($q = M_2 / M_1$) between 0.1 and 1, largely consistent with observational constraints \citep{sanaBinaryInteractionDominates2012,kobulnickyFreshCatchMassive2012,offnerOriginEvolutionMultiple2023}.
The initial separation, $a_\mathrm{i}$ is drawn from a flat-in-log distribution from \qtyrange{0.1}{1000}{au} \citep{abtNormalAbnormalBinary1983,sanaBinaryInteractionDominates2012}.
The eccentricity is set to zero at ZAMS.
We elaborate on our assumption on the CE parameters in Sect.~\ref{subsec:MT_and_CEE} and compact object remnants in Sect.~\ref{subsec:co}.

\subsubsection{Mass transfer and common envelope evolution}\label{subsec:MT_and_CEE}

MT is one of the most common ways through which stars in a binary can interact. MT occurs when the (donor) star fills its Roche lobe and begins to deposit mass onto its companion, the accretor star.
The point at which a donor star has filled its Roche lobe is determined using the fitting formulae for the Roche radii from \citet{eggletonAproximationsRadiiRoche1983}. The dynamical stability of MT, i.e., whether it results in stable RLOF or a CE event, is determined solely by the stellar properties at the onset of MT. Although the analytic prescriptions determining the stability are meant to capture the long term stability based on the instantaneous initial conditions, it is possible for stable MT to become unstable over time (e.g., \citealt{ivanovaCommonEnvelopeEvolution2020}). However, this is not currently implemented in \textsc{COMPAS}.

In order to assess the stability of the MT episode, we use the criterion of \citet{sobermanStabilityCriteriaMass1997}, which compares how the radius of the donor star changes with mass loss ($\zeta_\mathrm{star} \equiv \mathrm{d}\ln R_\mathrm{star}/\mathrm{d}\ln M$) versus how the radius of the Roche lobe changes with MT ($\zeta_\mathrm{RL} \equiv \mathrm{d}\ln R_\mathrm{RL}/\mathrm{d}\ln M$). 
This criteria is sensitive to the amount of mass accreted by the companion (accretor) and how much specific angular momentum is removed from the system by the non-accreted mass. If $\zeta_\mathrm{star}$ is greater than $\zeta_\mathrm{RL}$, then the MT is assumed to be stable.
The amount of mass removed from the donor depends on its stellar type.
For donors on the MS, mass is removed until they are able to fit within their Roche lobe, at which point the MT episode ends.
For donors with a defined core/envelope structure (i.e., post-MS), the whole envelope is removed on a thermal timescale.
The amount of accreted mass depends on the stellar type of the accretor.
In the synthetic population, we assume no accretion onto compact objects during CE evolution, as it does not impact the results \citep{grichenerMergersNeutronStars2023}.
Material that is not accreted is removed from the system with the specific angular momentum of the accretor (for additional details, see Sect.~4.2 of \citealt{team-compasCOMPASRapidBinary2022}).

Binary systems with NSs tend to experience dynamically unstable MT, which leads to CE evolution.
During CE evolution, the orbital separation can drastically decrease within several orbits; the CE evolution can result in the ejection of the envelope or merger.
In order to estimate the outcome of a CE episode, COMPAS uses the ``$\alpha_\mathrm{CE}$-$\lambda$" energy formalism to calculate the orbital separation following the CE phase \citep{webbinkDoubleWhiteDwarfs1984,livioCommonEnvelopePhase1988,ibenCommonEnvelopesBinary1993,ivanovaCommonEnvelopeEvolution2013}.
The energy formalism equates the change in orbital energy ($\Delta E_\mathrm{orb}$) before and after CE evolution with the binding energy of the CE ($E_\mathrm{bind}$) as
\begin{equation}\label{eqn:eformalist}
    E_\mathrm{bind} = \alpha_\mathrm{CE} \Delta E_\mathrm{orb},
\end{equation}
where $\alpha_\mathrm{CE}$ is the CE efficiency factor that parameterizes the fraction of orbital energy that can be used to eject the CE, which is set to unity in our simulations (i.e., $\alpha_\mathrm{CE}=1$).
COMPAS estimates the binding energy of the envelope by calculating the structure parameter $\lambda$ (on the order of unity) using the ``Nanjing lambda" prescriptions of \citet{xuBindingEnergyParameter2010,xuErratumBindingEnergy2010}, so that
\begin{equation}\label{eqn:ebind}
    E_\mathrm{bind} = -\frac{GMM_\mathrm{env}}{\lambda R},
\end{equation}
\noindent where $G$ is the gravitational constant, $M$ is the mass of the (secondary) star, $M_{\mathrm{env}}$ is the mass of the envelope, and $R$ is the stellar radius.
For our case of a NS engulfed by a star, we write Eqn.~\ref{eqn:eformalist} as

\begin{equation}\label{eqn:ebind_long}
    E_\mathrm{bind} = \alpha_\mathrm{CE}\left( \frac{GMM_\mathrm{NS}}{2a_\mathrm{pre-CE}} - \frac{GM_\mathrm{core}M_\mathrm{NS}}{2a_\mathrm{post-CE}} \right),
\end{equation}

\noindent where $M_\mathrm{core}$ is the core mass of the non-degenerate star at the onset of CE evolution, $M_\mathrm{NS}$ is the mass of the NS, and $a_\mathrm{pre-CE}$ and $a_\mathrm{post-CE}$ are the orbital separations before and after CE evolution, respectively. We can then solve for the post-CE separation by substituting Eqn.~\ref{eqn:ebind} for $E_\mathrm{bind}$ in Eqn.~\ref{eqn:ebind_long} in order to determine if a system merges. The binary is assumed to be circularized throughout the CE episode, and therefore the post-CE eccentricity is zero. 
In our synthetic population, we define a binary system as merged if it fits the condition $a_\mathrm{post-CE} \leq R_\mathrm{core}$. Contrary to \citet{grichenerMergersNeutronStars2023}, we take $R_\mathrm{core}$ as the radius of the stripped star as computed by \textsc{COMPAS}. If a star enters a CE with a NS companion and merges, we consider it to be a T\.ZO candidate.

\subsubsection{Compact object remnants}\label{subsec:co}

A SN must occur before the CE to form the NS that makes the core of the T\.ZO.
COMPAS considers four SN scenarios: regular core-collapse SNe (CCSNe), electron-capture SNe (ECSNe), ultra-stripped SNe (USSNe), and pair-instability SNe (PISNe). It uses the pre-SN core and envelope masses to determine SN type. Our synthetic population, at solar metallicity, does not contain any PISNe \citep{langerPairCreationSupernovae2007,farmerMindGapLocation2019}.

When a SN occurs, some material can fall back onto the core remnant. The fraction of material that falls back is known as the ``fallback fraction" and determines how much material is not ejected during the SN.
A fallback fraction of unity means that all material falls back onto the remnant, there is no ejecta and the stellar collapse is not observable. To determine the SNe rate ($N_\mathrm{SNe}$), we find all the CCSNe and ECSNe events with a fallback fraction less than unity. 
Out of the $10$ million simulated binaries, we find \num{4222994} core-collapse events predicted to be associated with an observable SN, and thus define $\mathcal{R} = N/N_\mathrm{SNe}= N/\num{4222994}$, which normalizes our event rates to the number of SNe (we do not consider Type Ia nor Type Ibc SNe in our estimates) in our synthetic population.
We only follow systems that remain bound after the first SN and we do not follow the evolution of unbound stars and DCOs. Therefore we do not account for the SN explosions that occur in these systems in our SN rate estimate. However, we estimate this to change the total SN rate by only a factor of $\lesssim 2$.

In our population, the maximum NS mass is set to \qty{2}{\msun} \citep{ozelBlackHoleMass2010}, which reproduces the hypothetical mass gap between NSs and BHs (although this is not confirmed, see, e.g., \citealt{siegelInvestigatingLowerMass2023} and \citealt{zhuFormationLowerMassgap2024}). The NS natal-kick distribution is bimodal Maxwellian, with CCSNe at the higher mode of \qty{265}{\kms} and ECSNe/USSNe at the lower mode of \qty{30}{\kms} \citep{vigna-gomezFormationHistoryGalactic2018}. The BH natal kicks are drawn from the same distributions and then the kick magnitude is reduced using the fallback model from \citet{fryerCompactRemnantMass2012}.

We assume \qty{100}{\percent} primordial binarity, although the previously mentioned \qty{1000}{au} upper limit on the semi-major axis means that approximately half of all binaries in the synthetic population do not interact at any point in their evolution.
If the binary has an eccentricity equal to or greater than unity after the natal kick, the binary becomes gravitationally unbound.

\subsection{Identifying subpopulations of interest}\label{subsec:subpops}
We filter our synthetic population in order to identify which systems have an evolutionary channel that leads to T\.ZO candidate formation.
Fig.~\ref{fig:evol_channel} shows an overview of the T\.ZO evolutionary channel studied in this paper, as explained in detail in Sect.~\ref{sec:intro}. Our goal is to characterize T\.ZO demographics, which involves also quantifying systems with similar evolutionary pathways.
To this effect, we first identify systems consisting of a star with a NS companion. We then distinguish between systems that later undergo CE evolution and those that become disrupted following the SN (the unbound stars). Systems that undergo CE evolution can result in either the ejection of their envelope (the stripped star and NS candidates, hereafter referred to as star + NS) or mergers of the NS with the stellar core (T\.ZO candidates).

As star + NS systems continue to evolve, the naked core can have its own core-collapse event (the second one in the system) and become a NS or a BH.
As previously shown, T\.ZOs and DCOs have similar evolutionary channels (see, e.g., \citealt{belczynskiFirstGravitationalwaveSource2016, vigna-gomezFormationHistoryGalactic2018, broekgaardenImpactMassiveBinary2021, grichenerMergersNeutronStars2023} and Fig.~\ref{fig:evol_channel}), so we also quantify the DCOs in the population, filtering as NS-NS, BH-BH, and NS-BH; and sub-categorizing by whether the binary will merge within the age of the Universe.
The coalescence time of DCOs via gravitational waves is based on \citet{petersGravitationalRadiationMotion1964}. If it is less than the Hubble time the system is a merger candidate.

Rejuvenated stars or their progenitors can be present in any of the aforementioned subpopulations, as rejuvenation only requires stable MT onto a MS star.
When rejuvenation occurs, the internal structure of the accretor is reorganized on a timescale comparable to a few convective turnover timescales of the accretor's core or the MT timescale, whichever is shorter. In most cases, this means a timescale comparable to or shorter than the global thermal timescale of the accretor \citep{renzoRejuvenatedAccretorsHave2023}.
As the thermal timescale of the accretor is much shorter than the typical remaining lifetime of the donor post-MT, it is likely that the accretor is in thermal equilibrium by the time the primary collapses and undergoes a SN.
We identify stellar companions in all of our subpopulations that had accreted via stable MT while on the MS and flag them as rejuvenated.

\subsection{Statistical uncertainties}
In order to assess the statistical uncertainty in the outcomes of our population synthesis simulations, we use the Clopper-Pearson method \citep{clopperUseConfidenceFiducial1934}. This approach, also known as the exact binomial confidence interval, is based on the cumulative distribution function of the binomial distribution. The statistical error arises from the finite sample size of 10 million stellar binary systems, within which we identify our different subpopulations.
If we were to draw different samples of 10 million systems, the numbers would vary due to random sampling variability. The Clopper-Pearson intervals account for this variability by providing a range within which the true results are likely to fall, with a specified confidence level. The intervals were computed using the \texttt{beta.ppf} function from the SciPy library, providing a robust measure of the statistical uncertainty associated with our estimated outcome rates. Within our synthetic population, we find the likelihood of having a statistically different number of T\.ZO progenitors, star + NS progenitors, or unbound stars under a \qty{1}{\percent} confidence level.

\subsection{Galactic rate estimates}\label{subsec:form_rates}
In order to estimate the number of T\.ZO candidates presently in the Milky Way, we require the time of T\.ZO formation, as well as the time the T\.ZO remains observable in that configuration, and the Galactic star formation rate (SFR).
For the duration in which T\.ZOs remain observable, i.e., the T\.ZO lifetime, we use the models from \citet{farmerObservationalPredictionsThorne2023} to approximate the T\.ZO lifetime as a function of the total T\.ZO mass.
Their models have a mass range of \qtyrange{5}{20}{\msun}, so we extrapolate the fit to be flat outside of this range. The resulting linear fit is

\begin{equation}\label{eqn:tzo_lifetime}
    \tau_\mathrm{T\dot{Z}O} = \left(9.29m + 7.15\right) \times \qty{e3}{yr},\quad 5\leq m/\unit{\msun} \leq 20,
\end{equation}

\noindent where $m$ is the mass of the T\.ZO in solar masses. \citet{farmerObservationalPredictionsThorne2023} determined a T\.ZO lifetime of $\tau_\mathrm{T\dot{Z}O} \approx 10^{4-5}$ years, which is comparable or shorter to models from \citet{cannonStructureEvolutionThorneZytkow1992,cannonMassiveThorneZytkowObjects1993}. We calculated our fit as accurate within $\lesssim$ \qty{6}{\percent}.

For the formation rate we follow the procedure outlined in Appendix B of \citet{chruslinskaDoubleNeutronStars2018}. In order to estimate when a T\.ZO progenitor is formed, we assume a constant SFR in the Galaxy, $\mathrm{SFR}_\mathrm{Gal} = 2.0 \pm 0.7 \, \unit{\msun\per \year}$ \citep{eliaStarFormationRate2022}. In practice, this means that each T\.ZO progenitor system is assigned a random birth time ($\tau_\mathrm{bin}$) between \qty{0}{Gyr} and \qty{10}{Gyr} ($T_\mathrm{Gal}$).
We then calculate the delay time for each T\.ZO ($\tau_\mathrm{delay}$), which is the sum of their binary progenitor birth time and their formation time ($\tau_\mathrm{form}$), $\tau_\mathrm{delay} = \tau_\mathrm{bin} + \tau_\mathrm{form}$.
We add the T\.ZO's lifetime from Eqn.~\ref{eqn:tzo_lifetime} to the delay time, $T = \tau_\mathrm{delay} + \tau_\mathrm{T\dot{Z}O}$. If this sum is greater than or equal to the age of the Galaxy, $T_\mathrm{Gal}$, then the T\.ZO currently exists and we increment $n_\mathrm{T\dot{Z}O}$, the unnormalized number of T\.ZOs in the Galaxy.

The final step to obtain estimations for the number of T\.ZOs in the Galaxy as well as T\.ZOs per solar mass is to normalize our values to the total stellar mass formed in the Galaxy over its lifetime.
We account for the full IMF, which includes stars less massive than the \qty{5}{\msun} lower limit in our synthetic population. We calculate a total stellar mass of $M_\mathrm{sim} = \qty{6e8}{\msun}$, assuming that all stars are created in binaries\footnote{As the upper limit for the initial period is \qty{1000}{au}, the population includes many non-interacting binaries that mimic stars that are born and evolve as single stars, see Sect.~\ref{subsec:co}.} and use a total mass range of \qtyrange{0.08}{100}{\msun}. The synthetic population uses a mass range of \qtyrange{5}{100}{\msun}, as a T\.ZO requires a NS to form and therefore progenitor systems must have a massive primary that experiences core collapse.
While stars with an initial mass of $\lesssim$\qty{5}{\msun} will not experience core-collapse, NS formation is still possible due to MT.
We then normalize $n_\mathrm{T\dot{Z}O}$ with $n_\mathrm{norm} = T_\mathrm{Gal} \times \mathrm{SFR}_\mathrm{Gal}/M_\mathrm{sim}$. The final number of T\.ZOs expected in the Galaxy is $N_\mathrm{T\dot{Z}O} = n_\mathrm{norm}n_\mathrm{T\dot{Z}O}$. T\.ZO formation rate per year is calculated as the raw number of T\.ZOs in our synthetic population divided by $M_\mathrm{sim}$ and multiplied by the SFR. We present the results in Sect.~\ref{subsec:formrates_results}.

\section{Results}\label{sec:results}

\subsection{Formation rates}\label{subsec:formrates_results}

\begin{figure}[!htbp]
	\centering
	\includegraphics{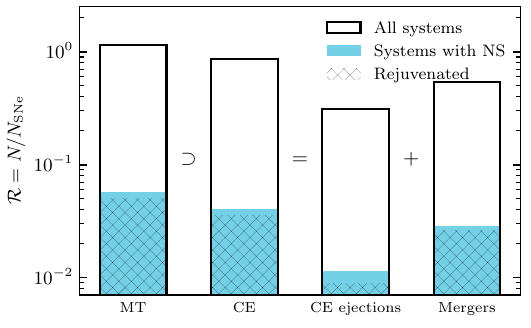}
	\caption{Summary of the event rates in our synthetic population at Solar metallicity \citep[$Z=0.0142$,][]{asplundChemicalCompositionSun2009}. We quantify the rate of binaries that experience the following (from left to right): mass transfer (MT), common envelope (CE) events (a subset of previous, can only lead to ejections or mergers), CE ejections, and mergers.
    For each column, the outline (black) indicates all binaries in the population that go through the event, the bar (light blue) indicates binaries with a NS, and hatching indicates companions to the NS rejuvenated by previous stable MT.
    The $y$-axis is in logarithmic scale and it shows the number of systems of interest ($N$) normalised by the number of SNe ($N_{\mathrm{SNe}}$, see Sect.~\ref{subsec:co} for more details).
    }
	\label{fig:rates_tzo}
\end{figure}

Here we present the formation rates of systems in the synthetic population (Fig.~\ref{fig:rates_tzo}).
Understanding the event rates in our population is key for constraining formation channels and predicting the likely end product of a binary.
We find that the rate of systems in the synthetic population that undergo MT is $\mathcal{R}\approx 1.2$, from which $\mathcal{R} \approx 0.85$ for systems that have a CE evolution phase.
There are more MT events than SNe, as the synthetic population has a lower mass limit of \qty{5}{\msun}, thus not all stars in the population will explode (see Sect.~\ref{subsec:co}).
While not all binaries in our synthetic population interact with their companions, those that do are likely to experience CE evolution.

Systems with a NS that undergo CE evolution are possible progenitors for either T\.ZOs or star + NS systems; in our synthetic population systems with a NS that undergo CE evolution have a formation rate of $\mathcal{R} \approx 0.040$ (second column of Fig.~\ref{fig:rates_tzo}). If we limit ourselves to systems with rejuvenated companions, the rate only drops by \qty{14}{\percent}, showing the importance of understanding the impact of rejuvenation on massive interacting binaries.

Systems with a NS that merge with the core are defined as potential T\.ZOs and we find a rate of $\mathcal{R}_\mathrm{T\dot{Z}O} \approx 0.029$. If the system does not merge, it is defined as a star + NS system and we find a rate of $\mathcal{R}_\mathrm{star + NS} \approx 0.011$.
We also estimate the number of T\.ZOs present in our Galaxy at the current moment (see Sect.~\ref{subsec:form_rates}).
We find a T\.ZO formation rate of \qty{\approx 4e-4}{\per \year} at solar metallicity. From this we estimate $\num{\approx 5}\pm 1$ T\.ZOs in the Milky Way at the present.

\begin{figure}[!htbp]
	\centering
	\includegraphics[width=\columnwidth]{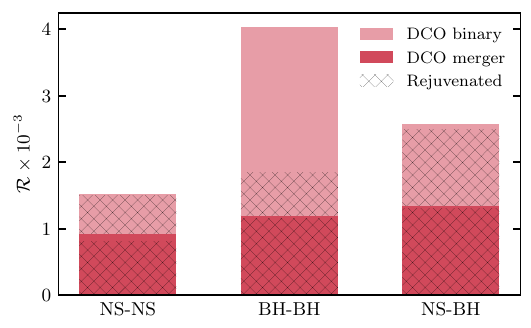}
	\caption{
    Summary of the double compact object (DCO) rates in our synthetic population at Solar metallicity ($Z=0.0142$). We quantify the rates of DCOs (from left to right): NS-NS binaries, BH-BH binaries, and NS-BH binaries. For each column, the upper portion (light pink) indicates non-merging DCOs, the lower column (dark pink) indicates DCOs that merge in a Hubble time, and the hatching indicates DCOs that had a rejuvenated companion before both SNe. 
    The $y$-axis is scaled as in Fig.~\ref{fig:rates_tzo}.
    }
	\label{fig:rates_dco}
\end{figure}

In our synthetic population, we find that a NS-NS binary is an alternative to T\.ZO formation (Fig.~\ref{fig:evol_channel}). However, due to the large uncertainties in MT prior to the second SN explosion and mass accretion onto the NS, we believe that NS-BH and BH-BH might also be an alternative outcome of the T\.ZO CE formation channel.
Fig.~\ref{fig:rates_dco} presents the DCO rates in our synthetic population, at solar metallicity.
These systems are categorized by whether they will merge within a Hubble time or remain a binary, with each bar in the figure representing the total number of systems of the stated type.
With the exception of BH-BH systems that do not merge, the DCOs in our population (mergers and not) are evenly split between staying a binary and merging due to gravitational wave emission.
NS-NS systems are the most rare, with $\mathcal{R}_\mathrm{NS-NS}\approx 0.002$, and the 
vast majority (\qty{91}{\percent}) of NS-NS systems had a rejuvenated companion before both SNe, regardless of whether or not they merge.
We find a BH-BH rate of $\mathcal{R}_\mathrm{BH-BH}\approx 0.004$,
with \qty{96}{\percent} of merging systems having had a rejuvenated companion before both SNe, and only \qty{23}{\percent} of non-merging systems with a rejuvenated companion before both SNe.
This is because merging systems must interact, increasing the chances of previous rejuvenation. Wider systems do not have this requirement and can evolve from MS to BH without ever interacting.
NS-BH systems have a rate of $\mathcal{R}_\mathrm{NSBH}\approx 0.003$ and \qty{96}{\percent} had a rejuvenated companion before the second SN.

We find that over \qty{90}{\percent} of DCO binaries have rejuvenated stars, indicating that including the effects of rejuvenation in COMPAS simulations and post-processing is pertinent.

\subsection{Progenitors in the Hertzsprung-Russell diagram}\label{subsec:HRD}

\begin{figure}[!htbp]
    \centering
    \includegraphics{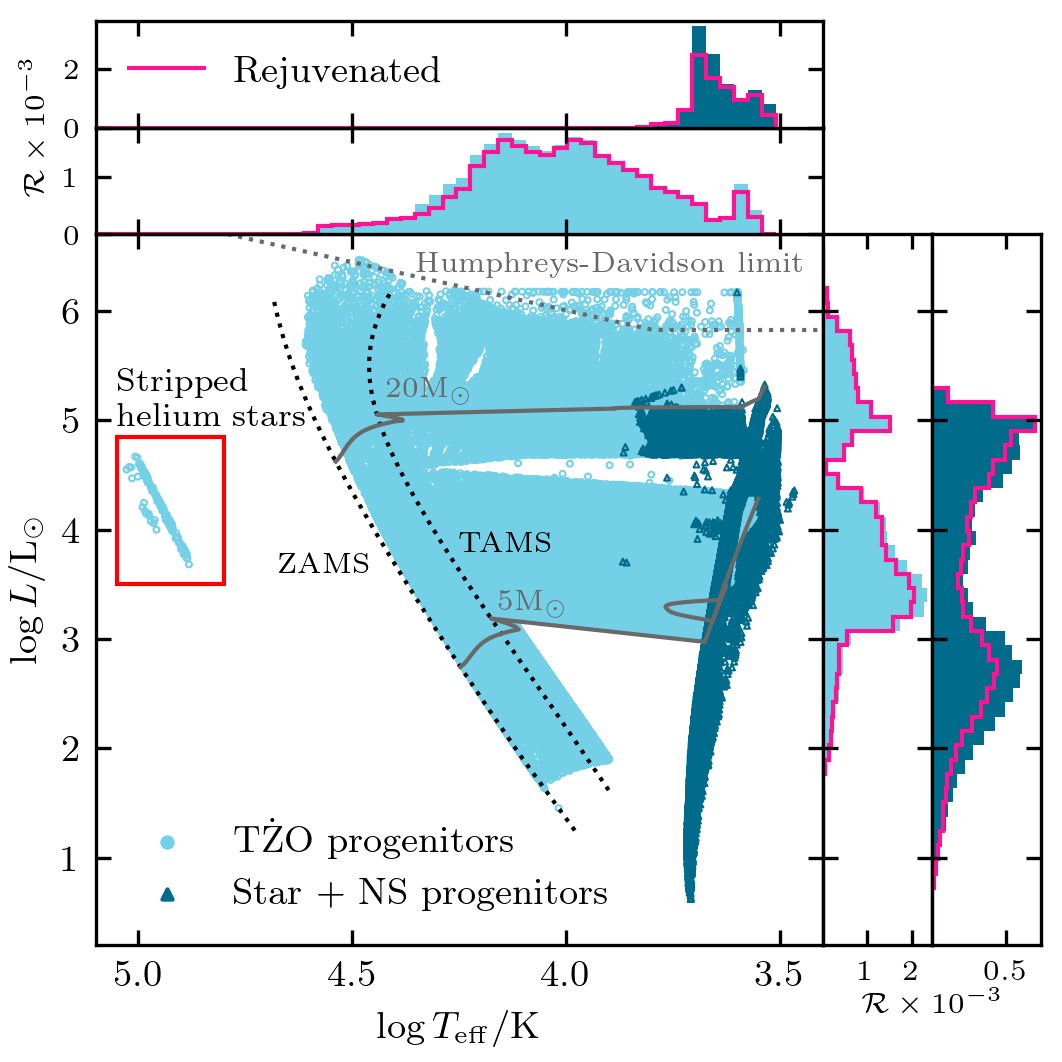}
    \caption{
    HRD showing the progenitors of T\.ZO candidates (light blue) and star + NS systems (dark blue) at Solar metallicity ($Z=0.0142$). 
    These progenitors are shown at the onset of RLOF leading to a CE event.
    Two representative single star tracks \citep{hurleyComprehensiveAnalyticFormulae2000} for \qtyrange{5}{20}{\msun} are shown as solid lines (gray).
    ZAMS and the terminal-age main sequence (TAMS) are shown as dotted lines (black). The Humphreys-Davidson limit \citep{HD_limit} is shown as a dotted line (gray).
    Above and to the right: histograms showing the rate density of T\.ZO and star + NS progenitors in their respective colors. 
    The outline (pink) indicates the subset of systems with a rejuvenated companion. 
    The $y$-axes of the histograms are scaled by the number of observable SNe in the synthetic population (see Sect.~\ref{subsec:co} for more details).
    }
    \label{fig:hrd_combine}
\end{figure}

Now that we have quantified the formation rates of our subpopulations of interest (Sect.~\ref{subsec:formrates_results}) we focus our attention on the location of T\.ZO and star + NS progenitor systems in the Hertzsprung-Russell diagram (HRD). Figure~\ref{fig:hrd_combine} presents systems with a star and a NS at the onset of RLOF leading to CE. They are further divided by whether they merge (T\.ZO candidate) or eject the CE (star + NS candidate). The histograms above and to the right of the main panel show that the majority of donors in both of these populations have experienced a previous stable MT event and become rejuvenated.

At the onset of RLOF, T\.ZO donor progenitors have a mass range of \qtyrange{\approx 2}{\approx 110}{\msun} and star + NS donor progenitors have a mass range of \qtyrange{\approx 2}{\approx 68}{\msun}.
Previous T\.ZO models have had an upper mass limit of \qty{\approx 60}{\msun} (e.g., \citealt{cannonStructureEvolutionThorneZytkow1992,farmerObservationalPredictionsThorne2023,hutchinson-smithRethinkingThorneZytkowObject2023}), meaning this population could potentially indicate T\.ZOs more massive than have ever been modeled.
We find that hotter stellar donors always result in mergers, while cooler donors might result in envelope ejection \citep[e.g.,][]{klenckiItHasBe2021}.
Examining the features of the HRD, we see stellar donors of T\.ZOs can be on the MS, even close to the ZAMS, when the CE begins.
Stellar donors for star + NS systems, however, are limited to the end of the Hertzsprung Gap (HG) and within the Asymptotic Giant Branch (AGB). There is a horizontal gap at $10^{4.5}$ \unit{L_\odot}, a numerical feature from the fitting formulae which enhances the gap between stars that begin core He burning while in the HG and stars that do not reach the required core temperature until after crossing the HG into the AGB (for more details, see \citealt{vigna-gomezCommonEnvelopeEpisodes2020}).
There is also a lack of T\.ZO progenitors past the MS for donors below \qty{5}{\msun}. These donor stars have white dwarf companions rather than NS companions. These systems may result in interesting transients, such as late SNe \citep{zapartasDelaytimeDistributionCorecollapse2017}.

\subsection{Kinematics}\label{subsec:kinematics}
\begin{figure}[!htbp]
	\centering
 	\includegraphics{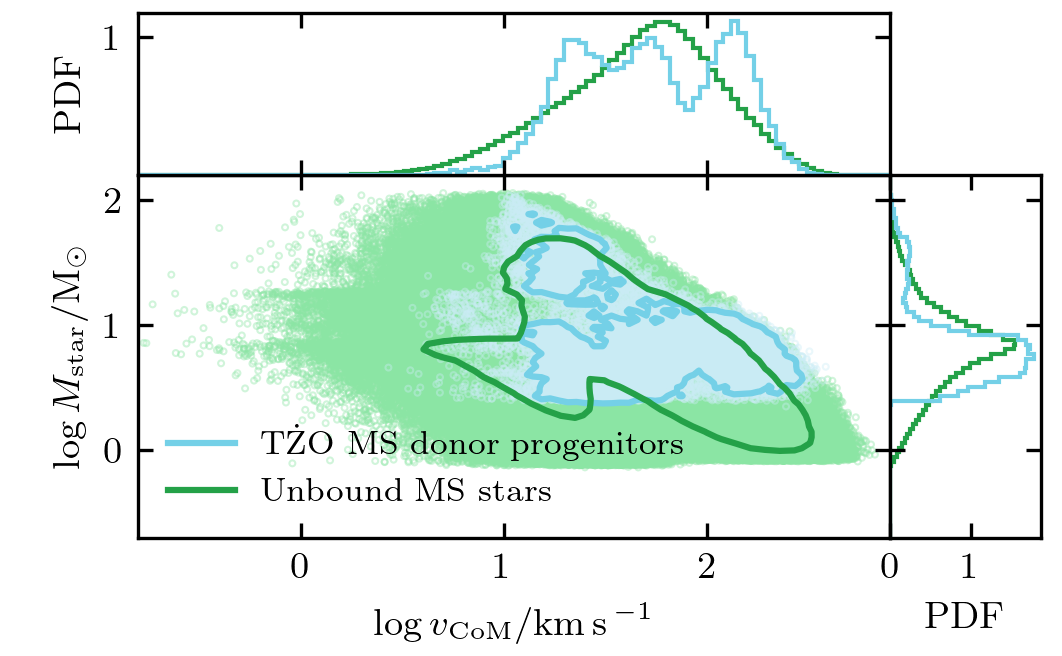}
 	\includegraphics{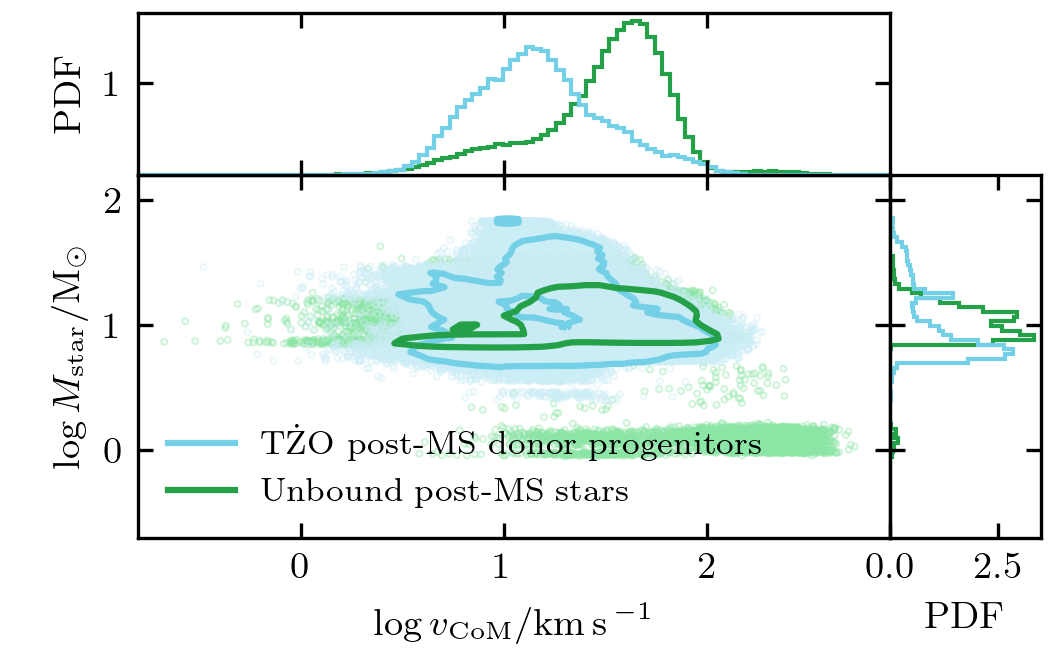}
   	\includegraphics{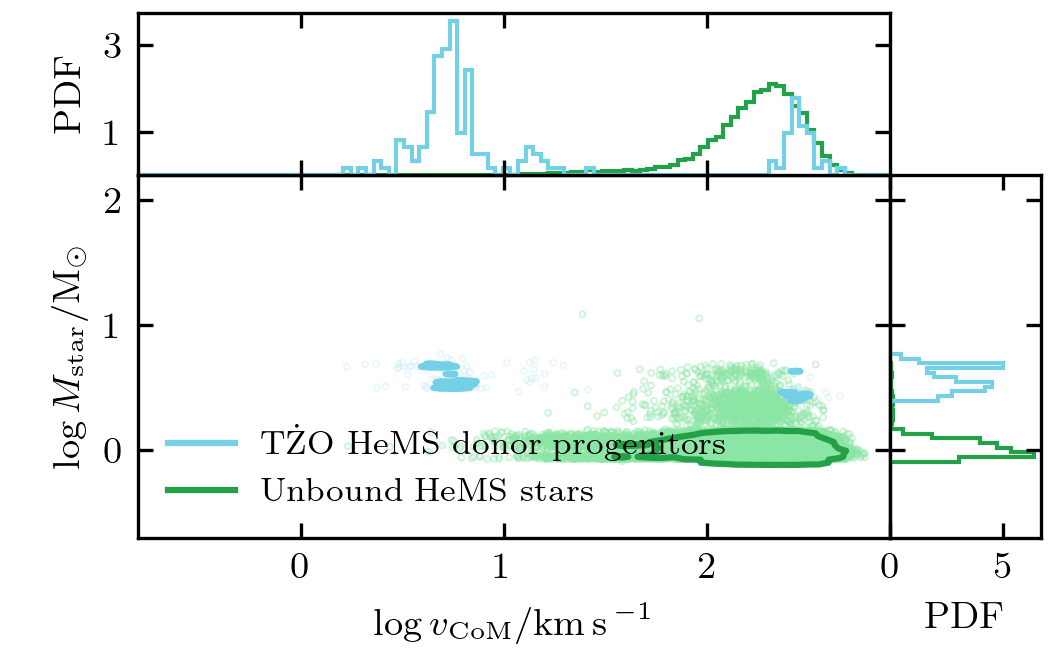}
	\caption{The mass of the secondary plotted against the center-of-mass speed for T\.ZO progenitors (light blue) and unbound stars (green) on the MS (top), post-MS (middle), and HeMS (bottom). The contours enclose \qty{90}{\percent} of the systems. Above and to the right of the main panel are histograms showing the density, normalized so that each integrates to unity.
    }
	\label{fig:sysspeed_corner}
\end{figure}

SN explosions lead to NS (and BH) formation, and the violent explosion is usually accompanied by a natal kick \citep[e.g.,][]{fryerCompactRemnantMass2012,vigna-gomezFormationHistoryGalactic2018}. The sudden change in the velocity of the newly born compact object alters the orbit and changes the center-of-mass velocity ($v_\mathrm{CoM}$) of the system. We investigate these center-of-mass "spatial velocities" as well as their relationship to the mass of the secondary (the donor) in Fig.~\ref{fig:sysspeed_corner}.
Here we explore if there are distinct, unequivocal signatures to differentiate between unbound stars (which become unbound after the first SN explosion and do not undergo CE evolution) and T\.ZOs.

Overall, the unbound stars (see step 4 of Fig.~\ref{fig:evol_channel}) dominate the parameter space, as the T\.ZO rate is $\mathcal{R}_\mathrm{T\dot{Z}O}\approx 0.029$ and the unbound star rate is $\mathcal{R}_\mathrm{unbound}\approx 0.83$, i.e. the formation rate of unbound stars in our synthetic population is approximately 27 times greater than that of the T\.ZOs.
It is useful to try and differentiate between T\.ZOs and unbound stars. We do so by comparing T\.ZO progenitors on the MS, past the MS, and at a stage where they become stripped stars depleted of hydrogen and rich in helium (HeMS), to unbound stars in the same respective stages.
In general, unbound stars are more massive and their velocity has negligible dependence on the natal kick of the NS or BH formed when the companion explodes, thus they move more slowly \citep{eldridgeRunawayStarsProgenitors2011,renzoMassiveRunawayWalkaway2019}. T\.ZOs are formed from binaries that do not become unbound when the primary undergoes a SN, meaning the magnitude of the natal kick of the NS is likely smaller (or it is in a tight binary), and $v_\mathrm{CoM}$ is strongly dependent on it \citep{eldridgeRunawayStarsProgenitors2011, renzoMassiveRunawayWalkaway2019, vandermeij:21}.

For the MS subpopulation (top panel of Fig.~\ref{fig:sysspeed_corner}), we find $\mathcal{R}_\mathrm{MS \: T\dot{Z}O} \approx 0.0071$ for the MS T\.ZO donor progenitors and $\mathcal{R}_\mathrm{MS \:unbound} \approx 0.71$ for the MS unbound stars.
The top panel of Fig.~\ref{fig:sysspeed_corner} shows nearly complete degeneracy in donor mass and $v_\mathrm{CoM}$.

For the subpopulation with stars beyond the MS (middle panel of Fig.~\ref{fig:sysspeed_corner}), we find $\mathcal{R}_\mathrm{post-MS \: T\dot{Z}O} \approx 0.022$ for the T\.ZO progenitors and $\mathcal{R}_\mathrm{post-MS \: unbound} \approx 0.045$ for the unbound stars.
The T\.ZO donor and unbound star progenitors show less overlap in both mass and $v_\mathrm{CoM}$. The T\.ZO progenitors have a higher mass range, reaching up to \qty{\approx 65}{\msun}, while the unbound stars have a maximum mass of \qty{\approx 30}{\msun}. The $v_\mathrm{CoM}$ of the T\.ZO progenitors suggest they move slower than the unbound stars, as higher velocities imply a higher chance of being disrupted. Mass, which is strongly correlated to luminosity, is a prominent distinguishing feature in these distributions; i.e., massive (greater than \qty{\approx 30}{\msun}), luminous objects might serve as an observable indicator for T\.ZOs. 

For the subpopulation with HeMS stars (bottom panel of Fig.~\ref{fig:sysspeed_corner}), we find $\mathcal{R}_\mathrm{HeT\dot{Z}O} \approx 4.0\times 10^{-5}$ for the HeMS T\.ZO donors (dashed arrow in Fig.~\ref{fig:evol_channel}) and $\mathcal{R}_\mathrm{HeMS \: unbound} \approx 0.008$ for the HeMS unbound stars. 
This subpopulation shows strong differentiation between the T\.ZO and unbound star progenitors.
While the number of HeMS T\.ZO donor progenitors in our synthetic population is two orders of magnitude smaller than that of the unbound stars, they have markedly higher donor masses, lower $v_\mathrm{CoM}$, and are depleted in hydrogen; these properties make them an extremely interesting future study prospect (see Sect.~\ref{disc_subsec:HeMSTZOS} for a discussion).

In total, we find that $v_\mathrm{CoM}$ and stellar mass are largely degenerate in the subpopulation of MS T\.ZO donor and unbound star progenitors, but are more clearly distinguished in the post-MS and HeMS subpopulations. The post-MS T\.ZO progenitors have a much higher maximum mass, which might serve as an observable indicator for T\.ZOs. The HeMS T\.ZO donor progenitors are small in number, but show clear differences to the unbound stars and are an interesting prospect for future studies. {As the unbound stars have a much higher rate than the T\.ZO progenitors, finding T\.ZOs among observations of massive stars is challenging. It is likely that to indisputably identify a T\.ZO would require a combination of an unusual abundance profile, kinematics, and perhaps stellar pulsations (e.g., \citealt{levesqueDiscoveryThorneZytkowObject2014,farmerObservationalPredictionsThorne2023}).} 

\section{Discussion}\label{sec:discuss}

We have presented population demographics of binary systems that enter a CE phase with a NS companion, using a solar metallicity-synthetic population generated with COMPAS and publicly available at Zenodo (see Sect.~\ref{subsec:data_initialdists}).
In the discussion, we focus on the subset of the population that is predicted to result in a merger during CE evolution, which are the potential progenitors of T\.ZOs.

\subsection{Rates}\label{disc_subsec:rates}

Throughout this work, we have focused on the formation rates of T\.ZO candidates, particularly CE evolution in field binaries under the premise that T\.ZOs can be successfully assembled through this channel (see Sect.~\ref{disc_subsec:sn_collision} for a discussion on formation from collisions following a SN).
Therefore, our quoted rates likely represent an optimistic estimate.
Under this premise, we find a T\.ZO formation rate per core-collapse SN of $\mathcal{R}_\mathrm{T\dot{Z}O} \approx 0.029$, which is equivalent to \qty{\approx 4e-4}{\per \year} in the Galaxy. From these values, we estimate \num{\approx 5\pm 1} T\.ZOs presently in the Milky Way (Sect.~\ref{subsec:formrates_results}). 
We now compare these rates with those in the literature.

\citet{grichenerMergersNeutronStars2023} also calculates rates of NS and stellar core mergers using the same population, although they exclude progenitor systems with a MS donor and calculate $R_\mathrm{core}$ using the fitting formulae of \cite{hallCoreRadiiCommonenvelope2014} for the $a<R_\mathrm{core}$ merger condition. Moreover, they normalize by number of core-collapse SNe in the synthetic population, not filtering by fallback fraction, and find a rate of $\approx 0.011$ relative to core-collapse SNe, approximately one third of our rate. As T\.ZO progenitors with MS donors make up \qty{\approx 25}{\percent} of our T\.ZO subpopulation and the SN filtering discussed in Sect.~\ref{subsec:co} takes out \qty{\approx 7}{\percent} of the SN population, our differing rates do not impact our conclusions.

\citet{cannonMassiveThorneZytkowObjects1993} estimates $10^7$ T\.ZOs in the history of the Galaxy. Using our value for the current number of T\.ZOs and the mean T\.ZO lifetime in our population (\qty{8.8e4}{yr}), we estimate \num{\approx 3e4} T\.ZOs in the history of the Galaxy. The large discrepancy is due to the fact that \citet{cannonMassiveThorneZytkowObjects1993} assumes that a T\.ZO is a later evolutionary stage for massive X-ray binaries (XRBs), with a significantly larger formation rate of \qty{\sim 0.001}{yr} \citep{meursNumberEvolvedEarlytype1989}. It is common for massive X-ray binaries to have a CE evolution phase, but not all CE evolution phases lead to mergers, which is illustrated by our subpopulation of star + NS systems (see Fig.~\ref{fig:rates_tzo}). 
However, the dichotomy between envelope ejection and stellar merger resulting in a T\.ZOs is not clearly understood, and will be dependent on the evolutionary history of the progenitor star (see Sect.~\ref{disc_subsec:progenitors} and \ref{disc_subsec:thin-envelope-TZOs}).

Our estimated formation rate matches the formation rates estimated by \citet{podsiadlowskiEvolutionFinalFate1995} and \citet{hutilukejiangFormationThorneZytkowObjects2018}.
They both consider the T\.ZO lifetime is \qtyrange{e5}{e6}{yr}, based on the calculations by \citet{cannonMassiveThorneZytkowObjects1993} and \citet{biehleObservationalProspectsMassive1994}.
Under these assumptions, they both predict similar rates: \citet{podsiadlowskiEvolutionFinalFate1995} predicts between $20$ and $200$ T\.ZOs in our Galaxy at any given moment, while \citet{hutilukejiangFormationThorneZytkowObjects2018} determines there are $15-150$ T\.ZOs.
In contrast, we use the \qtyrange{e4}{e5}{yr} estimated T\.ZO lifetime based on the models from \citet{farmerObservationalPredictionsThorne2023}, which results in only \num{5\pm 1} T\.ZOs presently in the Milky Way.
The major discrepancy between our number and the number found in \citet{podsiadlowskiEvolutionFinalFate1995} comes from the difference in assumed T\.ZO lifetime. \citet{farmerObservationalPredictionsThorne2023} analyzed the pulsational stability of T\.ZOs and found modes with growing amplitudes, which lead the models to shed their envelopes and stop being T\.ZOs, impacting the lifetime \citep[see also][]{Romagnolo_2024}.

\subsection{Formation uncertainties}

In this study, we assessed the role of recent work on stellar rejuvenation \citep{renzoRejuvenatedAccretorsHave2023}, core disruption during stellar merger events \citep{eversonRethinkingThorneYtkow2023,hutchinson-smithRethinkingThorneZytkowObject2023}, and T\.ZO lifetimes from stellar models \citep{farmerObservationalPredictionsThorne2023} on the predicted population of T\.ZOs at Solar metallicity.
However, we did not explore the many uncertainties from stellar and binary evolution in the proposed CE formation channel (Figure \ref{fig:evol_channel}), such as the initial distributions or the first SN leading to the NS formation, which inevitably propagates to the rate uncertainties for T\.ZOs.
Arguably, the main uncertainties are around the difficulties surrounding the CE phase.

The CE phase is complex and modeling its evolution is challenging \citep{ivanovaCommonEnvelopeEvolution2013, ropkeSimulationsCommonenvelopeEvolution2023}, making the determination of the outcome of a CE episode a rather uncertain endeavor.
In rapid population synthesis, many simplifying assumptions are made in order to determine which systems engage in a CE phase and, if they do, which systems lead to a CE (partial or total) ejection, or alternatively result in a merger.
In order to determine which systems engage in a CE phase, COMPAS relies on mass-radius exponents which determine how the radius of the donor star responds to stripping of its envelope (see Sect.~\ref{subsec:MT_and_CEE}).
For giant stars, which we consider as likely leading to a CE phase, COMPAS defines the mass-radius exponents in terms of polytropes \citep{sobermanStabilityCriteriaMass1997}.
However, polytropes are not the only way to calculate the mass-radius exponents. They can also be calculated with detailed one-dimensional (1D) stellar models, as has been done by e.g., \citet{geAdiabaticMassLossIII2020} and \citet{2023A&A...669A..45T}, which tentatively suggests that MT is more stable than previously thought.
These models incorporate very rapid mass transfer and can account for processes like the delayed dynamical instability, which is not currently included in COMPAS \citep[e.g.,][]{geAdiabaticMassLossI2010}.
While 1D detailed models give important insights, however, it is important to note that three-dimensional radiation hydrodynamic simulations of the convective envelopes of giants suggest that 1D models are likely failing to accurately approximate radiation transport and opacities \citep{chiavassaThreedimensionalHydrodynamicalSimulations2010,jiangThreeDimensionalNatures2023,goldbergNumericalSimulationsConvective2022}, which directly affects the stellar radius.

Other methods for determining whether MT leads to a CE episode depend on either a stellar-phase-dependent critical mass ratio \citep[e.g.,][see Sect.~\ref{disc_subsec:xrbs}]{claeysBinaryProgenitorModels2011} or a threshold for the formation of a convective envelope in the donor \citep{2024ApJ...969....1P,Romagnolo_2024}, for which RLOF is then assumed to result in a CE phase.
However, stability of a MT episode with a NS accretor is not only determined by the structure and response to mass loss of the donor, but also by the response of the accretor to mass gain \citep[e.g.,][]{kippenhahnRadiiAccretingMain1977, weiEvolutionFinalFate2023, 2024ApJ...966L...7L} and is also very sensitive to assumptions about how the non-accreted mass is lost from the binary
\citep[e.g.,][]{willcoxImpactAngularMomentum2023,geAdiabaticMassLossV2024}.

\citet{willcoxImpactAngularMomentum2023} used COMPAS to study the impact of updating these criteria with new prescriptions for the critical mass ratios and therefore MT stability based on \citet{claeysBinaryProgenitorModels2011} and \citet{geAdiabaticMassLossIII2020}, finding that the mass-radius exponents are generally higher for all masses and radii, signifying that unstable MT (and therefore CE evolution) may occur less frequently than previously thought. 
These prescriptions assume fully conservative MT, which might not be the case for the T\.ZOs progenitors.
In this work, we explore T\.ZOs which generally experience two MT episodes (Figure \ref{fig:evol_channel}).
The first MT episode, in which rejuvenation occurs, involves a donor star with a less evolved MS companion and mass ratio close to that at ZAMS; it is generally uncertain to what extent this MT episode is conservative.
In the second MT episode, which potentially leads to T\.ZO formation, there is a donor star with a NS companion and a more unequal mass ratio; this MT episode is likely highly non-conservative.
\citet{geAdiabaticMassLossV2024} explored non-conservative MT, particularly in the context of NS high-mass X-ray binaries.
They find that dynamically unstable mass transfer is more likely to occur for systems with extreme mass ratios and wide orbital periods, which correspond to more luminous and more extended systems in our population (Figure \ref{fig:hrd_combine}). For a deeper discussion of uncertainties in T\.ZO formation, see \citet{ogradyThorneZytkowObjects2024}.

\subsection{Progenitors}\label{disc_subsec:progenitors}
From our synthetic population, we determine: i) in which binary star systems there is MT from a non-degenerate star onto a NS companion, ii) if these MT episodes are dynamically unstable and lead to CE evolution, and iii) if CE evolution will ultimately lead to a merger and thus possibly the formation of a T\.ZO.
There are several uncertainties in the stellar evolution leading to the onset of MT (RLOF) and arguably even more in the merger process and the potential assembly of a T\.ZO \citep{langerPresupernovaEvolutionMassive2012,pavlovskiiMassTransferGiant2015,eversonRethinkingThorneYtkow2023,ivanovaUnifiedRapidMass2024}.
In this paper, we focus on providing insights in the properties of the T\.ZO stellar progenitor rather than on the T\.ZOs themselves.
Understanding the evolutionary pathways of T\.ZOs, and providing their more common configurations will be crucial in order to assess whether or not a T\.ZO can form (see Sect.~\ref{disc_subsec:thin-envelope-TZOs}).

Generally, the non-stripped stellar progenitors of T\.ZOs can be anywhere between the ZAMS and the (super)giant phase, with masses between \qty{2}{\msun} and \qty{110}{\msun}, and effectively anywhere in the HRD (Fig.~\ref{fig:hrd_combine}). There is, however, a slight preference for less evolved and more massive hot donors with envelopes that are significantly more gravitationally bound than those of their less massive cool counterparts. 
In our models, there is a non-negligible fraction ($\approx$\qty{15}{\percent}) of progenitors with masses above \qty{20}{\msun} (middle panel of Fig.~\ref{fig:sysspeed_corner}), which we consider as potential candidates for T\.ZOs (but see Sect.~\ref{subsec:kinematics} for a discussion about unbound stars).

We find that the vast majority (\qty{92}{\percent}) of T\.ZO progenitors have donor stars which have been rejuvenated. This means that the donor star has previously stably accreted mass during a MT episode (Fig.~\ref{fig:rates_tzo}). Here, we analyzed the synthetic solar metallicity population from \cite{grichenerMergersNeutronStars2023}. Metallicity influences the stellar radii across the evolution through different physical effects \citep[e.g.,][and references therein]{xin:23}. The possible variation with metallicity of the fraction of rejuvenated donors for CE with a NS remains to be studied.

Rejuvenation alters the structure of the accretor star, especially at the core-envelope boundary (see Sect.~\ref{sec:intro} for more details). The structural changes are highly uncertain and are likely to vary depending on the evolutionary stage of the star when rejuvenation occurs. \citet{renzoRejuvenatedAccretorsHave2023} propose that the altered structure could lower the binding energy of the envelope.
Under the energy formalism used throughout this work (Sect.~\ref{subsec:MT_and_CEE}), the binding energy of the envelope must be lesser than $\alpha_\mathrm{CE} \Delta E_\mathrm{orb}$ for envelope ejection to occur.
In our T\.ZO progenitor population, $E_\mathrm{bind}$, while large enough to prevent envelope ejection, was less than \num{1.5} times $\alpha_\mathrm{CE}\Delta E_\mathrm{orb}$. Assessing the impact of structural changes in rejuvenated stars will require 3D simulations \citep{landriTheEffectofDonorStarRejuv2024}.

A decrease in the binding energy of the envelope of the donor could potentially have catastrophic effects on the formation rate of T\.ZOs due to the substantial increase in systems where the envelope is unbound, but would increase the formation of binaries with a NS and a stripped star. If we assume that T\.ZOs with rejuvenated stellar donors would be unable to form through the CE evolution channel due to the decrease in binding energy, the stripped binary formation rate could increase by a factor of four (Fig.~\ref{fig:rates_tzo}). For these stripped binaries, the non-degenerate massive star in the binary would continue to evolve and potentially form a compact object. If the objects are close enough, they could merge within the age of the Universe, emitting gravitational waves detectable by ground-based observatories. Alternatively, the formation of wide DCO binaries in the Galaxy could also be detectable by space-based gravitational wave detectors (see \citealt{lauDetectingDoubleNeutron2020}, \citealt{waggGravitationalWaveSources2022}). In summary, rejuvenation could potentially increase DCO formation and merger rates in the isolated binary formation scenario (but only by a factor of a few), and could drastically decrease the T\.ZO formation rate.

\subsection{X-ray binaries as progenitors}\label{disc_subsec:xrbs}

Constraining T\.ZO rates using observations is difficult, as we are yet to confirm one and they are predicted to look very similar to red supergiants \citep{thorneRedGiantsSupergiants1975,thorneStarsDegenerateNeutron1977,levesqueDiscoveryThorneZytkowObject2014,farmerObservationalPredictionsThorne2023}.
Previous estimations of the T\.ZO rate has been based on forward modeling of XRBs \citep[e.g.,][]{podsiadlowskiEvolutionFinalFate1995}, as T\.ZOs and high-mass XRBs share an evolutionary path.
Here we briefly comment on population studies of these systems in order to gain some insights on their observed and predicted rates.

Unlike their less massive counterparts, high-mass XRBs transfer their mass through stellar winds. 
Because of their relatively short orbital periods ($\sim 1-10$ d), most of these high-mass XRBs will eventually expand and potentially fill their Roche lobes.
Nevertheless, the outcome of the Roche lobe overflow episode is uncertain.
Using a population of \num{69} Be XRBs from the \citet{coeCatalogueBeXray2015} catalog, \citet{vinciguerraBeXrayBinaries2020} argues that in order for a system to result in a merger, massive donors need to be extended and have an extreme mass ratio. In another study, \citet{hutchinson-smithRethinkingThorneZytkowObject2023} investigated LMC X-4, a tight (semi-major axis $\qty{\approx 14.2}{R_\odot}$) star + NS binary in the Large Magellanic Cloud with a mass ratio of \num{0.09} that has some of the most accurately measured orbital parameters of high-mass XRBs \citep{falangaEphemerisOrbitalDecay2015}.
They predict that after filling its Roche lobe, the stellar companion will engulf the NS, which will merge with the core, accrete mass, and subsequently collapse into a BH and form a thin-envelope T\.ZO (see Sect.~\ref{disc_subsec:thin-envelope-TZOs}). It is not unreasonable to expect similar systems in the Milky Way.

\citet{geAdiabaticMassLossV2024} calculated critical mass ratios for non-conservative mass transfer for stellar models from \qtyrange{1}{100}{\msun} at $Z = 0.001$.
They fit the maximum critical mass ratio as a function of the orbital period and found that as orbital period increased, the required mass ratio for unstable MT becomes more extreme, with a minimum of $q = 0.05$ at $P_\mathrm{orb} = \qty{400}{d}$. Even for short period binaries ($\sim$\qty{1}{d}), they predict a maximum mass ratio of \num{0.4}, above which MT is unstable and results in a CE episode. Finally, they compare 14 NS high-mass XRBs to their orbital period-mass ratio fit and find that 10 of them lie in the unstable MT parameter space, although they speculate that 7 of them might be helium stars, which have different critical mass ratios (see Sect.~\ref{disc_subsec:HeMSTZOS}). We used the T\.ZO progenitors in our synthetic population that have mass ratios and periods within the parameter space of the \citet{geAdiabaticMassLossV2024} study to do a rough comparison to the mass ratio-orbital period fit. We found that \qty{\approx 70}{\percent} fit the criteria for unstable MT, meaning that a CE would still form. A T\.ZO is still not certain at this step, as the binary must merge inside the envelope rather than eject it. Studies of the impact of the common envelope efficiency parameter on envelope ejection rates for star + NS binaries show that with the most extreme assumptions the number of mergers decreases by an order of magnitude \citep{grichenerMergersNeutronStars2023}.

\subsection{Core-disruption and thin-envelope T\.ZOs}\label{disc_subsec:thin-envelope-TZOs}
\begin{figure}[!htbp]
	\centering
	\includegraphics[width=\columnwidth]{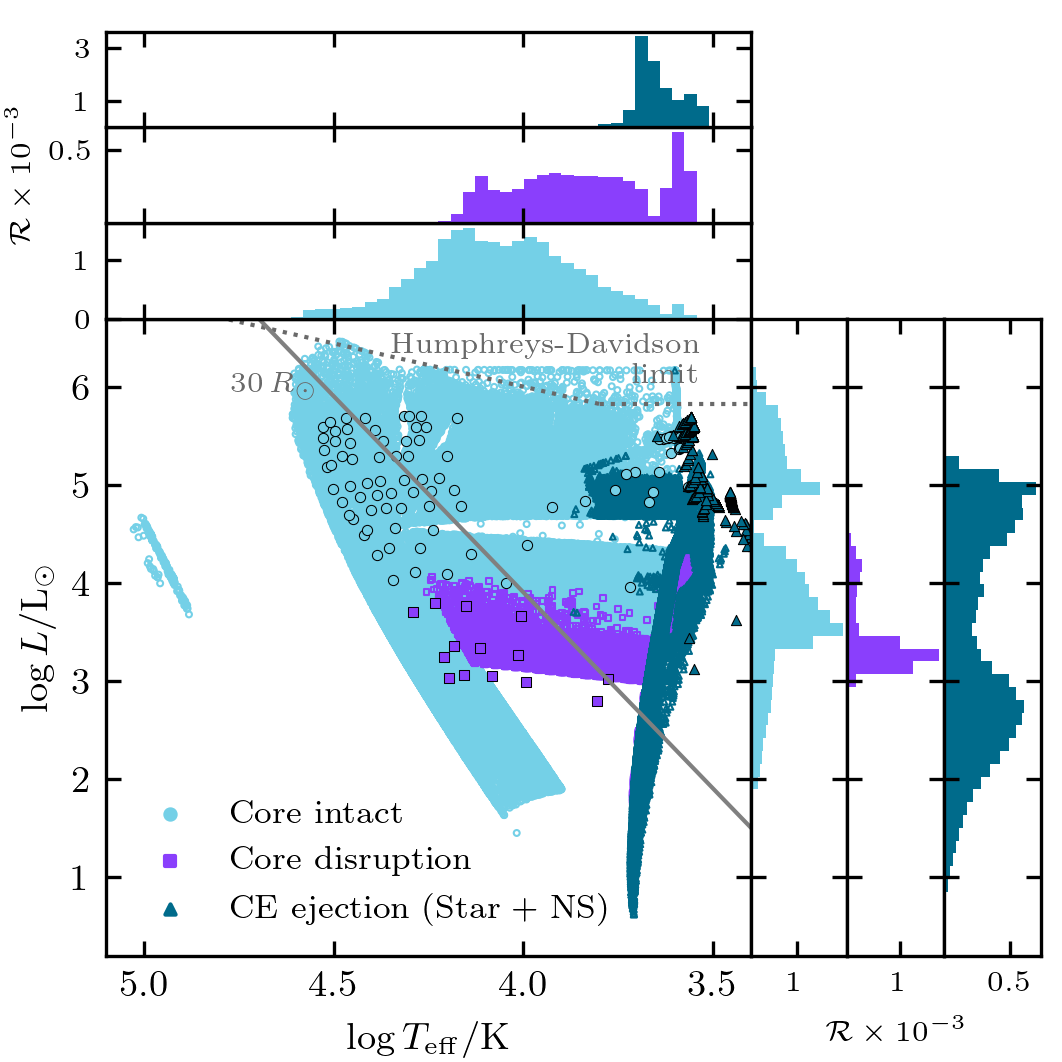}
	\caption{
    HRD showing an alternative outcome of the CE phase (cf. Fig.~\ref{fig:hrd_combine}).
    We analyze our synthetic population in the context of the evolutionary models from \cite{eversonRethinkingThorneYtkow2023}. 
    We assess the outcome of the CE evolution based on the criteria described by \cite{eversonRethinkingThorneYtkow2023} and show systems predicted to have an intact core (light blue), disrupted core (purple), and those that eject the CE (dark blue). The data from \citet{eversonRethinkingThorneYtkow2023} are overplotted as the same symbols with black outlines.
    We show a \qty{30}{\rsun} isoradial line, the cutoff for TET\.ZO formation, as a solid line (gray).
    The $y$-axes of the histograms are scaled by the number of observable SNe in the synthetic population (see Sect.~\ref{subsec:co} for more details).}
	\label{fig:TETZOs}
\end{figure}

Recent studies by \citet{eversonRethinkingThorneYtkow2023} and \cite{hutchinson-smithRethinkingThorneZytkowObject2023} have once again called into question the possibility of forming T\.ZOs \citep[see also, e.g.,][]{toutHV2112ThorneZytkowObject2014,maccaroneLargeProperMotion2016,beasorCriticalReevaluationThorneZytkow2018}, specifically in field binaries undergoing CE evolution. 
In their 3D hydrodynamical merger simulations, shocks generated by the inspiraling compact object propagate through the companion star and spin up the envelope and core, such that conditions are met to form a central accretion disk upon completion of the merger. Accretion disk formation disallows the internal structure required to form a stable T\.ZO \citep{podsiadlowskiEvolutionFinalFate1995}, so the parameter space study of \citet{eversonRethinkingThorneYtkow2023} sought to determine cases in which accretion disk formation might be avoided, first looking at populations in which the stellar core of the companion remains intact through the merger.
They provide a criterion for core disruption based on the tidal radius of the system ($r_{\rm tidal} \approx R_{\rm core}\, q_{\rm c}^{1/3}$; $q_\mathrm{c} \equiv M_\mathrm{NS}/M_\mathrm{core}$), such that the core will be destroyed and form an accretion disk about the NS upon merger if $r_{\rm tidal} > R_{\rm core}$.
This translates to a simple metric based on the mass ratio, $q_\mathrm{c} \leq 1$, for the core to remain intact.
Applying this metric, we find that \qty{\approx 20}{\percent} of T\.ZO progenitors in our synthetic population satisfy this criterion for core disruption (Fig.~\ref{fig:TETZOs}).
These systems are preferentially low-luminosity ($\lesssim 10^{4} \, \unit{L/L_\odot}$) post-MS stars.
More luminous progenitors have cores massive enough that they generally avoid core disruption; this is primarily due to the binding energy of the core being directly proportional to the square of mass.

Using our synthetic population, we find that a non-negligible fraction of T\.ZO progenitors are more massive and luminous, covering a wider temperature range than those modeled by \citet{eversonRethinkingThorneYtkow2023}.
Their models are exclusively post-MS with an upper mass limit of $\lesssim \qty{40}{\msun}$, lacking the resolution to coarsely fill the crossing of the HG.

A caveat within our method with respect to the simulations of \citet{eversonRethinkingThorneYtkow2023} is that our stellar tracks are based on fitting formulae \citep{hurleyComprehensiveAnalyticFormulae2000} which do not contain the details of the structure of the core. These simplifying assumptions mean that we can only estimate $q_\mathrm{c}$ and thus the outcome of NS plunge-in.
One effect of these differences can be seen in Fig.~\ref{fig:TETZOs}, where the less luminous MS progenitors in our model retain an intact core when detailed stellar models in hydrodynamic simulations clearly suggest they will experience core disruption.
For MS donors, we assume the core mass of the star to be approximately equal to its total mass, which naturally results in an overestimation of the core mass. Taking a fraction of the star's mass as the core would increase the $q_\mathrm{c}$ favoring an outcome of core disruption.

Finally, \citet{eversonRethinkingThorneYtkow2023} find that all of their models, with both disrupted and non-disrupted cores, will form an accretion disk due to the angular momentum deposited by the NS during inspiral and merger, therefore preventing the assembly of stable T\.ZOs.
From these failed T\.ZOs, they suggest instead that a thin-envelope T\.ZO (TET\.ZO) may be formed after the NS plunges in and subsequently collapses into a BH due to mass accretion. For these systems, if the binding energy of the core released during collapse or as accretion feedback is not sufficient to fully eject the post-merger remnant envelope, a small fraction ($\lesssim 1 \%$) of the initial envelope mass may remain as an optically-thick thin-envelope around the BH. The thin-envelope reprocesses accretion luminosity from the central engine and emits in the X-ray band, possibly contributing to the population of ultraluminous X-ray sources.
We explore their metrics within our population and provide an estimate of the number of T\.ZO progenitors which could become possible TET\.ZOs.
We estimate the formation rate of these TET\.ZOs by considering the progenitors in which the companion radius is $\leq\qty{30}{\rsun}$ (below which TET\.ZO formation may be most likely; see Fig.~7 of \citealt{eversonRethinkingThorneYtkow2023}) and estimate their lifetime following Eqn.~13 from \citet{eversonRethinkingThorneYtkow2023}.
From our population, and following the metrics provided by \citet{eversonRethinkingThorneYtkow2023}, we obtain \num{\approx 1.1e-4} TET\.ZOs per \unit{\msun}, which corresponds to $\mathcal{R}_\mathrm{TET\dot{Z}O} \approx 0.015$ and results in \num{1.5\pm 0.5} TET\.ZOs in the Milky Way, which would form instead of the \num{5\pm 1} T\.ZOs previously discussed.

\subsection{Helium-rich T\.ZOs}\label{disc_subsec:HeMSTZOS}
We find a small fraction (\qty{\approx 0.1}{\percent} of our T\.ZO subpopulation) of systems in which the donor is a HeMS star (see Sect.~\ref{subsec:kinematics}).
These HeMS donors originate via the same formation pathway discussed throughout this work. Following Fig.~\ref{fig:evol_channel}: after the successful ejection (step 6) of the CE, the newly stripped HeMS star in the binary (step 7) might later fill its Roche lobe (step 8), beginning a new MT episode.
Typically, this episode of MT proceeds in a dynamically stable manner \citep{hjellmingThresholdsRapidMass1987,sobermanStabilityCriteriaMass1997,vigna-gomezFormationHistoryGalactic2018}. The CE evolution leading to HeT\.ZOs occurs \qtyrange{\sim e4}{\sim e6}{\year} after the binary's first CE event, well after the stripped star has had time to thermally readjust \citep{2022MNRAS.511.2326V}.
In our population, \qty{\approx 8}{\percent} of the HeMS stars with a NS companion experience dynamically unstable MT, resulting in a HeT\.ZO.
We find a formation rate of $\mathcal{R}_\mathrm{HeT\dot{Z}O}\approx \num{4.0e-5}$ for these systems, which may constitute a lower limit for the following reasons.
First, the radii of these stars may be underestimated, although the effect is small at high metallicity, and stability criteria in this regime are highly uncertain and untested \citep{laplaceExpansionStrippedenvelopeStars2020}.
Second, COMPAS does not account for tidal evolution nor circumbinary disks in post-CE binaries \citep[see, e.g.,][and references therein]{2024A&A...688A.128V} that could lower the separation of post-CE systems, increasing the number of mergers. Therefore, both the possible underestimation of the radius and the effect of tides could mean we are underestimating the number of HeT\.ZOs.

HeMS T\.ZO progenitor donor stars in our synthetic population have luminosities between $\approx 5\times 10^3$ \unit{\lsun} and $\approx 5 \times 10^{4}$ \unit{\lsun} with masses in the range between \qtyrange{\approx 2.5}{\approx 6}{\msun}.
With respect to their parameters as star + NS binaries, the mass ratio ($M_{\rm{HeMS}}/M_{\rm{NS}}$) of these systems is from \numrange{2.3}{5.3} and the $\zeta_\mathrm{RL}$ is between approximately \numrange{2}{8}.
In COMPAS, the radius-mass exponent of HeMS stars is simplistically defined as a fixed number ($\zeta_{\rm{HeMS}}=2$) and the accretion rate onto a NS is determined by the Eddington limit \citep{team-compasCOMPASRapidBinary2022}, which is why the condition for dynamical instability ($\zeta_{\rm{RL}}>\zeta_{\rm{HeMS}}$) is met for a fraction of systems.
It is not clear if these configurations should actually result in a CE phase.
\cite{hjellmingThresholdsRapidMass1987} used polytropic models to assess rapid MT in binaries; for radiative MS stars, they do find that a dynamical instability might arise if the mass ratio is above the critical value of 2.14.
Another approach is to consider the structure and response of HeMS stars, which is done by \citet{zhangAdiabaticMassLoss2024} and \citet{geAdiabaticMassLossV2024}. \citet{zhangAdiabaticMassLoss2024} assumes fully conservative MT and finds the critical mass ratio below which unstable MT occurs lies between $0.38$ and unity. \qty{65}{\percent} of our HeMS T\.ZO progenitor donors have a mass ratio below $0.38$, indicating that many of these systems would still enter into a CE.
Conversely, \citet{geAdiabaticMassLossV2024} assumes non-conservative MT and calculates a fit for the mass ratio-orbital period space for the He stellar donors (at $Z = 0.02$), finding a maximum mass ratio of $0.1$ for unstable MT. If we follow this criteria, all of our HeMS T\.ZO donor progenitors undergo stable MT, meaning no CE and therefore no formation of HeMS T\.ZOs.

In Fig.~\ref{fig:sysspeed_corner} we show that HeMS T\.ZO systems have markedly different position in the mass and $v_\mathrm{CoM}$ parameter space with respect to unbound HeMS stars. If formed, T\.ZOs from the HeMS evolutionary channel might be identifiable through their slow spatial velocities, unusually high masses, and lack of hydrogen spectral lines. These donor stars have a mass range of \qtyrange{2.5}{5.9}{\msun}, which is distinguishable from their unbound counterparts. Even if rare, these objects form an interesting avenue for future study, for example in the context of Luminous Fast Blue Optical Transients (\citealt{metzgerLuminousFastBlue2022,sokerCommonEnvelopeJets2022}).

\subsection{Nucleosynthetic signatures}\label{disc_subsec:nucleo}
While the potential nucleosynthetic signature of high mass T\.ZOs has been discussed (see \citealt{levesqueDiscoveryThorneZytkowObject2014}), the impact of rejuvenation and HeMS donors on the nucleosynthetic signature has not.
In the case of rejuvenation, the accreted material is rich in He and N and poor in O and C, but these differences would not impact the outcome of the irp-process.
Therefore, we do not expect nucleosynthesis from T\.ZOs to be significantly different due to previous rejuvenation.

Helium content, however, could have a noticeable impact on the abundance profile.
The irp-process in high mass T\.ZOs is expected to lead to production of rubidium, strontium, yttrium, and molybdenum \citep{cannonMassiveThorneZytkowObjects1993} as well as an overabundance of lithium (e.g., \citealt{podsiadlowskiEvolutionFinalFate1995}).
These spectral features have been key tools in the search for T\.ZO candidates \citep{levesqueDiscoveryThorneZytkowObject2014,ogradyCoolLuminousHighly2020,ogradyCoolLuminousHighly2023}.
However, \citet{farmerObservationalPredictionsThorne2023} find that evolutionary models with very high initial helium fractions can lack Rb.
Our synthetic population predicts that a small fraction of T\.ZOs with high initial helium fractions will form, in particular those with post-terminal age MS or HeMS donors (\qty{\approx 0.5}{\percent} of our T\.ZO subpopulation).
These T\.ZOs would likely have a distinct abundance profile, differing from the T\.ZOs previously considered in the literature \citep{farmerObservationalPredictionsThorne2023}.

\subsection{Collisions following a Supernova} \label{disc_subsec:sn_collision}
While not the main focus of this paper, T\.ZOs have also been predicted to form via collisions following a SN (SN-T\.ZOs).
In this channel, the SN kick propels the newly-formed NS directly into the companion star, which could potentially form a T\.ZO.
We calculated the rate of systems in our synthetic population where one of the stars has experienced a SN resulting in a NS, and the companion is a non-degenerate star, as $\mathcal{R} \approx 0.032$.
We find that \qty{\approx 1}{\percent} of these systems have a periastron $\leq$\qty{1}{\rsun} directly after the SN explosion and thus could possibly result in a T\.ZO candidate from the SN collision scenario; this order-of-magnitude estimate results in a formation rate of $\mathcal{R}_\mathrm{SN-T\dot{Z}O} \approx \num{3.4e-4}$.
While this rate is low, these could still become relevant to the T\.ZO population and are worth consideration.

\citet{leonardNewWayMake1994} estimated a Galactic formation rate for this channel of \qty{0.25}{kpc^{-2}\, Myr ^{-1}} for T\.ZOs with an envelope mass over \qty{5}{\msun}. Using a Galactic radius of \qty{15}{kpc} and extrapolating their rate to the Galactic disk gives \qty{2e-4}{\per \year}.
Our previously computed formation rate of \qty{\approx 4e-4}{\per \year} is higher than the above extrapolated rate.
However, \citet{leonardNewWayMake1994} only considers dynamically-formed T\.ZOs, and notes that including the CE evolution channel could increase the formation rate.

\citet{podsiadlowskiEvolutionFinalFate1995} proposed that the SN colliding channel could be of comparable importance to the CE channel, the latter with an estimated rate of \qty{\sim e-4}{\per \year}. \citet{renzoMassiveRunawayWalkaway2019} optimistically estimated that \num{e-4} T\.ZOs would form per binary SN events at the time of the SN forming the NS, based on the solid angle seen by the companion.
Starting with the rate of binaries where one is a NS and the other is a non-degenerate star, using the formation rate of \num{\sim e-4} per core-collapse events in a binary from \citet{renzoMassiveRunawayWalkaway2019} results in $\mathcal{R} \approx \num{3e-6}$ T\.ZOs formed via collision post-SN in our synthetic population.
Translating this into a rate per year (using the previously mentioned SFR), we obtain \qty{e-4}{\per \year}, the same as \citet{podsiadlowskiEvolutionFinalFate1995} estimated for the collision channel and lower than the rate estimated with the CE channel.
Additionally, \citet{hiraiNeutronStarsColliding2022} studied the scenario in which the NS becomes unbound and then collides with its companion, but concluded that such events were rare enough to not impact previous estimates of the T\.ZO rate.

\section{Summary}\label{sec:summary}
In this study, we analyzed the Galactic metallicity, CE efficiency parameter at unity synthetic population from \citet{grichenerMergersNeutronStars2023} with a focus on potential progenitors of T\.ZOs assembled in the field via CE evolution and SN kick-induced mergers.
We used this data to explore the potential progenitors of T\.ZOs assembled in the field, particularly in the context of the CE channel.
The T\.ZO donor progenitors found in this population have an upper mass limit of \qty{\approx 110}{\msun}, which is significantly more massive than existing T\.ZO models and suggest that T\.ZOs should be explored at higher mass ranges.

\paragraph{Rates.}
We found a local T\.ZO formation rate of \qty{\approx 4e-4}{\per \year} or $\mathcal{R}_\mathrm{T\dot{Z}O}=N_\mathrm{T\dot{Z}O}/N_\mathrm{SNe} \approx 0.03$, in agreement with other studies in the literature (e.g., \citealt{cannonMassiveThorneZytkowObjects1993,podsiadlowskiEvolutionFinalFate1995,hutilukejiangFormationThorneZytkowObjects2018}).
We used T\.ZO lifetimes as calculated in the 1D detailed stellar simulations from \cite{farmerObservationalPredictionsThorne2023} which are of order \qtyrange{e4}{e5}{\year}.
We predicted \num{\approx5\pm 1} T\.ZOs presently in the Milky Way.
We found that our formation rates are overall consistent with those in the literature and that the predicted number of Galactic T\.ZOs is largely sensitive to the T\.ZO predicted lifetime.
\paragraph{Rejuvenation.}
We found that \qty{92}{\percent} (\qty{78}{\percent}) of T\.ZO (star + NS) progenitor stellar donors have previously accreted mass during a MT episode and have become rejuvenated. 
If the lower binding energy of rejuvenated stars results in easier CE ejection, we speculate that many of the T\.ZO progenitors in our population could actually result in star + NS systems, and potentially decrease (increase) the formation rate of T\.ZOs (DCO formations and mergers).

\paragraph{Core disruption and TET\.ZOs.}
We used the criterion of \cite{eversonRethinkingThorneYtkow2023} and found that the majority of our T\.ZO progenitors avoid core disruption.
We compared our results to the T\.ZO models from \cite{eversonRethinkingThorneYtkow2023}, which suggest that excess angular momentum deposited during CE evolution likely prevents the formation of canonical T\.ZOs in field binaries, identifying unexplored regions of the parameter space described by our synthetic population. \citet{eversonRethinkingThorneYtkow2023} present TET\.ZOs as a hypothetical merger product and X-ray source that may form from failed T\.ZOs; we estimated the number of TET\.ZOs in the Milky Way to be \num{1.5\pm 0.5}.

\paragraph{He-rich T\.ZOs.}
We found a rare outcome ($\mathcal{R}_\mathrm{HeT\dot{Z}O}\approx 4\times 10^{-5}$) in which a Helium-rich, stripped progenitor merges with a NS companion, a configuration that is generally not accounted for in the literature. These configurations would likely emit a distinct spectral signature, and therefore pose an interesting topic and merit further study.

\paragraph{Kinematics.}
We compared T\.ZO progenitors and unbound stars in mass and spatial velocity and found nearly complete degeneracy for progenitors stars on the MS.
However, post-MS and HeMS T\.ZO progenitor stars are more massive and have lower spatial velocities than their unbound counterparts.
We find that T\.ZO progenitors can be difficult to distinguish from stars which become unbound after the first SN, while also being significantly less common.

\paragraph{SN colliding T\.ZOs.}
We briefly explored T\.ZOs that form when the SN kicks the newly formed NS into the companion star. We found these systems and filtered for those with a periastron $\leq \qty{1}{\rsun}$ directly after the SN explosion in order to estimate a formation rate of $\mathcal{R}_\mathrm{SN-T\dot{Z}O} \sim \num{e-4}$. We consider that it is possible that these systems collided and merged into T\.ZOs.

\begin{acknowledgements}
We thank Martyna Chruslinska, Selma de Mink, Ryosuke Hirai, Cole Johnston, Stephen Justham, Jing-Ze Ma, Ilya Mandel, Ruggero Valli, and Reinhold Willcox for useful discussions. 
We thank Richard O'Shaughnessy for his support.
We thank the Kavli Foundation and the Max Planck Institute for Astrophysics for supporting the 2023 Kavli Summer Program during which much of this work was completed. 
KN thanks the LSST-DA Data Science Fellowship Program, which is funded by LSST-DA, the Brinson Foundation, and the Moore Foundation; her participation in the program has benefited this work. 
AVG acknowledges funding from the Netherlands Organisation for Scientific Research (NWO), as part of the Vidi research program BinWaves (project number 639.042.728, PI: de Mink).
AG acknowledges support from the Miriam and Aaron Gutwirth Fellowship and from the Gruber Fellowship funded by the IAU and The Gruber Foundation.
RWE acknowledges support from the Heising-Simons Foundation.
\end{acknowledgements}

\bibliographystyle{aa}
\bibliography{refs}




\label{lastpage}
\end{document}